# Electron energy loss spectra of $Na_{0.33}CoO_2 \cdot yH_2O$ (y = 0, 0.6 and 1.3)

J.Q. Li, H.X. Yang, R.J. Xiao, Y.G. Shi and H.R Zhang

Beijing National Laboratory for Condensed Matter Physics, Institute of Physics, Chinese Academy of Sciences, Beijing 100080, China

Electron energy loss (EEL) spectra have been obtained on materials with the nominal compositions of $Na_{0.33}CoO_2 \cdot yH_2O$ (y = 0, 0.6 and 1.3). Spectral analyses revealed systematic changes in the low-loss energy region (< 10eV) and the core losses (O-K, $Co-L_{3,2}$) along with the water intercalation. The inter (intra)-band transitions and the relevant plasmon resonance were found to be sensitive to water content. Variations of the electronic structure and the Co valence states during hydration have been discussed based on our experimental results.

**Tel: +86-10-82649524; Fax: +86-10-82649531**

**E-mail address: Ljq@ssc.iphy.ac.cn (Jianqi Li)**

$Na_{0.33}CoO_2$ and the related materials are receiving great interest due to the discovery of superconductivity in the hydrated material of $Na_{0.33}CoO_2 \cdot yH_2O$ (y ~ 1.3) in addition to the known anomalous thermoelectric properties [1-4]. These materials, similar to the high-$T_c$ cuprates, have the layered structure consisting of the two dimensional $CoO_2$ sheets and intercalated $Na^+$ ions. Recent investigations suggest that electron correlations between Co 3d electrons are substantial for the presence of these significant properties in present system [5-7]. Singh explained this kind of notable thermoelectric effect in terms of the large density of states near the Fermi surface [8]; whereas some others [6, 7] proposed that the large thermopower at room temperature arises from the strong correlation of 3d electrons. Recent analyses of the triangular cobalt oxides $Na_xCoO_2$ with large thermopower coefficient reveals the existence of characters similar to the resonating valence bond (RVB) phase [4, 5]. Actually, understanding the electronic states and structures of Co 3d and O 2p bands in the correlated electron scenario is of the central importance for gaining insight into the anomalous properties in $Na_xCoO_2 \cdot yH_2O$ materials. The $Na_{0.33}CoO_2 \cdot yH_2O$ system in general contains three relatively stable phases with different water contents [9, 10, 11], i.e. the metallic $Na_{0.33}CoO_2$ phase normally deintercalated of Na from $Na_{0.75}CoO_2$, the intermediate $Na_{0.33}CoO_2 \cdot 0.6H_2O$ phase and the superconducting $Na_{0.33}CoO_2 \cdot 1.3H_2O$ phase. In this paper, we report on the EEL spectroscopy analysis of $Na_{0.33}CoO_2 \cdot yH_2O$ (y = 0, 0.6 and 1.3) materials. We have observed numerous features that can be attributed to the collective plasmon excitations, interband transitions, or core-level transitions. We have also found certain remarkable changes in EEL spectral structures along with the water intercalation.

Both single crystalline $Na_xCoO_2$ (0.3 < x < 0.75) samples and polycrystalline of $Na_{0.33}CoO_2 \cdot yH_2O$ (y = 0, 0.6 and 1.3) samples have been used in the present study. The sample syntheses and characterizations have been reported in ref. [12, 13]. Specimens for transmission-electron microscopy (TEM) observations were prepared by gently crushing the powder material into fine fragments, which were then supported by a copper grid coated with a thin carbon film. Thin single crystalline samples were obtained by peeling them from large single crystals with a tape, and then mounting them on standard electron microscopy grids. The EEL spectroscopy measurements were performed on a Tecnai F20 transmission electron microscope equipped with a post column Gatan imaging filter. The energy resolution in the EEL spectra is 0.75 eV under normal operation conditions. Our EELS experiments were performed with the convergence angle of ~0.7 mrad and the spectrometer collection angle of ~3.0 mrad. It is noted that the $Na_{0.33}CoO_2 \cdot yH_2O$ materials are easily damaged under electron beam due to the high mobility of Na atoms as well as the weak bonding of $H_2O$ molecules inside the crystals; we therefore performed all TEM observations at the low temperature of around 100K. It is found that radiation damage can be almost eliminated during our measurements.

Figs.1 show the electron energy loss spectra in the low energy range obtained from samples with nominal compositions of $Na_{0.33}CoO_2$, $Na_{0.33}CoO_2 \cdot 0.6H_2O$ and $Na_{0.33}CoO_2 \cdot 1.3H_2O$, respectively. Energy resolution in these spectra is about 0.9 eV for the FWHM (full width at the half maximum) of the zero-loss peak. The slight asymmetry of the elastic zero-loss peaks is, in fact, due to the presence of low energy excitations. Each spectrum in general contains four notable peaks labeled as *a* through *d*. Two peaks from the transitions of the localized Na and Co core states can be easily identified at the positions of 33.2 eV (peak *c*) for Na-L loss and 63 eV (peak *d*) for Co-M loss. The broad large peaks at around 25 eV (peak *b*) in all three spectra are due to the collective plasmon excitations of all the valence electrons. These highly damped bulk plasmon peaks are easy to identify due to their large excitation cross sections for fast electrons. We have performed a brief estimation on the collective plasmon excitations with Drude formula of $\hbar\omega_p = \hbar(ne^2/\varepsilon_0 m)^{1/2}$ for the free electron approximation, and get $\hbar\omega_p = 21.5$ eV for $Na_{0.33}CoO_2$. The density of valence electron n in this material is about $3.4 \times 10^{29}\,m^{-3}$. The discrepancy between theoretical result and the experimental data of $25 \pm 0.3$ eV is proposed to be chiefly in connection with the binding strength of the valence electrons with the crystal lattice [14]. The most remarkable phenomenon revealed in our experiments is the apparent changes in the spectral features below 10eV; both peak position and intensity vary systematically along with the water intercalation. For instance, the strong peak at 8 eV (peak *a*) in $Na_{0.33}CoO_2$ shifts to lower energy of ~6.9 eV in $Na_{0.33}CoO_2 \cdot 0.6H_2O$, and then to 6.0 eV in the $Na_{0.33}CoO_2 \cdot 1.3H_2O$ superconductor. This strong plasmon peak is considered to arise mainly from a collective excitation in connection with the interband transitions from nonolocalized Co-O2p state just below Fermi level $E_f$, to the empty sates just above $E_f$. Careful examinations suggest that all EEL spectra also contain many fine features in addition to these strong excitations. Especially, the spectrum from the superconducting phase $Na_{0.33}CoO_2 \cdot 1.3H_2O$ reveals the presence of several clear small peaks/shoulders below 10 eV.

The theoretical and experimental studies of electronic structure in this kind of materials have been reported in numerous literatures [15-17]. As a result of the two-dimensional nature in both the hydrated and nonhydrated compounds, the energy bands dispersion in $Na_{0.33}CoO_2$ and $Na_{0.33}CoO_2 \cdot 1.3H_2O$ along the *z*-direction is small. The Co 3d bands are crystal field split in the octahedral O environment. Two narrow bands, derived respectively from $e_g$ and $t_{2g}$ states, exist near the Fermi energy, the trigonal symmetry of the Co site splits the $t_{2g}$ state further into the $a_{1g}$ state and the doubly degenerate $e_{2g}$ states. The O2p bands extend from approximately -7 eV to -2 eV, relative to the Fermi energy $E_f$ . They are clearly separated from the transition metal *d* bands. The Co3d–O2p hybridization is weak. Fig. 2 is a brief energy level scheme for $Na_{0.33}CoO_2 \cdot yH_2O$, qualitatively illustrating the occupied and unoccupied states from $-20$ eV below $E_f$ and 15 eV above $E_f$. It should be mentioned that the specific role of the intercalated water on superconductivity is still a open question, electronic structure calculations demonstrated that the presence of additional water layers can make this system more two dimensional, but no evident alternation appears on the electronic structure near the fermi surface [17].

Fig. 3 shows the EEL spectrum in the low-loss region ($<$ 20eV) for the superconducting $Na_{0.33}CoO_2 \cdot 1.3H_2O$ and metallic $Na_{0.33}CoO_2$ samples, demonstrating the presence of small peaks and shoulder in addition to the bulk plasmon excitations as indicated by small vertical lines. We have carefully analyzed the spectra obtained from different samples to identify the noteworthy peaks based on the band structure as mentioned above. The excitation involving core levels are usually easy to identify in accordance with their characteristic peak energies. We can thus identify the 13 eV peak in the $Na_{0.33}CoO_2 \cdot 1.3H_2O$ spectrum as the excitation associated with H1s (hydrogen K edge); the small kink at 19 eV as the O2s core excitation (oxygen $L_1$ loss); the other small prominent peaks are likely to be interband or intraband transitions. Since the states near $E_f$ are derived from Co3d and O2p orbitals, we therefore identify the 4eV peak to a transition from the extended Co3d state just below $E_f$ to the states in the conduction band just above Fermi level (from $t_{2g}$ to $e_g$). The 8.8 eV peak is assigned to the extended O2p states to the conduction band (O2p to $e_g$). In the $Na_{0.33}CoO_2$ spectrum, carefully examination suggest there are indeed several weak or poorly resolved shoulders, the 5.8 eV shoulder can be assigned to an interband transition of O2p to $a_{1g}$ and the shoulder at 19 eV is a core loss peak from the O2s excitation. In order to facilitate the comparison, we have listed the main observed peaks from all $Na_{0.33}CoO_2 \cdot yH_2O$ (y = 0, 0.6 and 1.3) samples in table 1, and assigned the peaks based on the band scheme shown in Fig. 2. Our systematic results directly demonstrate that the energy excitations below 10eV change evidently with water intercalation, therefore, it is possible that the electronic structure of the layered system has certain changes near the Fermi level during hydration. This alternations possibly arise from band splitting in connection with interlayer coupling, the $CoO_2$ layers are more isolated in the hydrated cases, and thus more two dimensional, this feature is commonly believed to be an important ingredient for the occurrence of anomalous properties in this type of materials as broadly discussed in the study of high Tc superconductors [18].

Fig. 4 shows the dielectric functions obtained by Kramer-Kronig analysis (KKA) of the spectrum in the low-energy loss region. We followed the standard procedure by subtracting the contributions from the elastic peak, then deconvoluting to remove the contribution due to multiple scattering by applying Fourier-log method, resulting in the single scattering distribution. Before the KKA transformation can be carried out, the energy-loss function $Im(-1/\varepsilon(0))$ has to be suitably extrapolated to cover the whole energy range. The extended energy-loss function can then be normalized by taking the real part of the inverse of the dielectric function $Re(1/\varepsilon(0))$ to be zero, as is appropriate for metal. It is noted that the dielectric functions for $Na_{0.33}CoO_2$ and $Na_{0.33}CoO_2 \cdot 1.3H_2O$ have very similar structure in the shown energy range. The small peaks appear in the imaginary part $\varepsilon_2$ are considered to be the interband transitions. At the low energy end, the energy-loss function is overwhelmed by the elastic peak which has to be subtracted from the spectra. This introduces considerable uncertainty at energy below 3.5 eV under our experimental conditions. In order to obtain the relatively precise optical conductivity and reflectivity below 5 eV, further studies are still under progress.

Fig. 5a shows the oxygen K-edge core loss EEL spectra after background subtraction for $Na_{0.33}CoO_2 \cdot 1.3H_2O$, and $Na_{0.33}CoO_2$ materials. Five peaks (**a** to **e**) ranging from 520 to 580 eV are observed on each spectrum. The peaks (labeled **a)** correspond to the transition from O1s towards Co3d-O2p hybridized vacant states, it should be predominated by the $a_{1g}$ and $e_g$ bands as illustrated in Fig. 2. It is noted that the relative weight of this contribution, as compared to the total intensity of the O K edge, decreases evidently in the water intercalated superconducting $Na_{0.33}CoO_2 \cdot 1.3H_2O$. This reflects a decrease of accessible vacant Co-3d states along with hydration. Peak *c*, at around 540 eV, reflects a transition from O1s to O2p state hybridized with the more delocalized transition metal Co-4s and H-1s states. The structural feature of this peak can be understood based on intrashell multiple scattering within six-oxygen coordination shell as reported in ref. 19. Peak *d* lying to the high-energy-loss side of peak *c* and peak *e* at 570 K arise dominantly from single-scattering events from different oxygen coordination shells as pointed out by H. Kurata et al [19]. The most striking feature in Fig. 5a is the evident growth of peak *b* in the superconducting sample in comparison with that of $Na_{0.33}CoO_2$, this feature is considered arising from the excitations of O1s core level of $H_2O$ intercalated among $CO_2$ layers, these results are fundamentally in good agreement with the electronic structure which showing the H1s empty state ranges from 5 eV to 12 eV above the Fermi level (Fig. 2) in $Na_{0.33}CoO_2 \cdot 1.3H_2O$. High energy synchrotron photoemission spectroscopy measurements on the $Na_{0.33}CoO_2 \cdot 1.3H_2O$ samples reveals the presence of a sharp peak at 533 eV which was attributed to O1s core level of water [20], this peak position is slightly lower than that of our observations. Actually, we have performed several experiments on a variety of samples with different superconducting $T_c$, the major water O-**K** loss in all samples appears at around 536 eV. In view of the fact that synchrotron photoemission spectroscopy is sensitive to the surface states, this discrepancy therefore possibly arises from the difference between surface and bulk hydrated states. Figure 5b shows the EEL spectrum of Co **L₃** and **L₂** edges of $Na_{0.33}CoO_2 \cdot 1.3H_2O$ and $Na_{0.33}CoO_2$ obtained at the temperature of 100 K. These spectra (so called white lines) exhibit main peaks derived from Co $2p_{3/2}$ and $2p_{1/2}$ due to spin-orbit splitting. The intensities of **$L_3$** and **$L_2$** white lines are related to the unoccupied states in the 3d bands. In $Na_{0.33}CoO_2 \cdot yH_2O$ (y = 0, 0.6 and 1.3) materials, we have made a series of measurements on the intensity ratio of $L_3/L_2$ and analyzed by the method as reported by Wang et al [21]. The ratio $L_3/L_2$ is around 2.2, these data could yield the Co valence of about 3.6 for the sample of $Na_{0.33}CoO_2$. Careful examinations of spectra from the hydrated samples suggest that the $L_3/L_2$ rises slowly with water intercalation. This fact suggests that the valence state of Co in the superconducting sample is possibly lower than that of the $Na_{0.33}CoO_2$. This result is also in agreement with the results reported in ref. 22. The exact estimation of the valence state from EEL spectra is very difficult; we must consider the local magnetic moments, core hole effects and other factors. Analysis of the oxygen K loss spectra was also performed in correlation with the alternation of Co valence state during water intercalation. It can be clearly recognized that the 528 eV peaks (see fig. 5a peak *a*) is considerably stronger in $Na_{0.33}CoO_2$ than that in $Na_{0.33}CoO_2 \cdot 1.3H_2O$. This feature directly indicates that water intercalation could lower the valence state of Co ions in present system [23, 24].    .

In conclusion, measurements of EEL spectra from $Na_{0.33}CoO_2 \cdot yH_2O$ (y = 0, 0.6 and 1.3) revealed a rich variety of properties of electronic structure for this kind of layered system. Evident changes have been observed in EEL spectra of the low energy range (< 10eV) and the high-energy core losses along with water intercalation. Interband transitions and plasmon extractions have been carefully analyzed in several typical samples, in particular the $Na_{0.33}CoO_2 \cdot 1.3H_2O$ superconductor. The electronic structure and the Co valence states are likely to vary progressively along with hydration.

**Acknowledgments**

We would like to thank Prof. N.L. Wang for providing single crystal samples of $Na_xCoO_2$ and Dr. Q. Li for very valuable discussions. The work reported here is supported by National Natural Foundation of China.

Figure captions

Fig.1 The electron energy loss spectra in the low energy range from samples with nominal compositions of $Na_{0.33}CoO_2$, $Na_{0.33}CoO_2 \cdot 0.6H_2O$, and $Na_{0.33}CoO_2 \cdot 1.3H_2O$, respectively. All spectra are taken slightly off the <001> direction.

Fig. 2 A brief energy level scheme for $Na_{0.33}CoO_2 \cdot yH_2O$, qualitatively illustrating the occupied and unoccupied states from $-20eV$ below $E_f$ and $15eV$ above $E_f$.

Fig. 3 Electron-energy-loss spectra of $Na_{0.33}CoO_2$ and $Na_{0.33}CoO_2 \cdot 1.3H_2O$ in the range from 0 to 20eV, illustrating the presence of a series of weak peaks.

Fig. 4. Dielectric functions of $Na_{0.33}CoO_2$ and $Na_{0.33}CoO_2 \cdot 1.3H_2O$.

Fig. 5 Electron-energy-loss spectra of $Na_{0.33}CoO_2$ and $Na_{0.33}CoO_2 \cdot 1.3H_2O$ in the energy regions of (a) oxygen **K** absorption edge and (b) Co **L** absorption edge.

Table 1. The peak positions and brief descriptions of the nature of the transitions in EEL spectra obtained from $Na_{0.33}CoO_2 \cdot yH_2O$ (y=0, 0.6, and 1.3). The uncertainty in the energy positions is partial given in parentheses

| $Na_{0.33}CoO_2$ | | $Na_{0.33}CoO_2 \cdot 0.6H_2O$ | | $Na_{0.33}CoO_2 \cdot 1.3H_2O$ $T_c$=4.6K | |
|---|---|---|---|---|---|
| Peak(eV) | Description | Peak(eV) | Description | Peak(eV) | Description |
| 5.8 | O2p to $a_{1g}$ | 5.5($\pm$0.2) | O2p to $a_{1g}$ | 4.0 | Co $t_{2g}$ to eg |
| 7.9 ($\pm$0.3) | Plasmon | 7($\pm$0.3) | Plasmon | 6($\pm$0.3) | Plasmon |
| | | | | 8.8 | O2p to $e_g$ |
| | | | | 13.8 | H-K |
| 19 | O-$L_1$ | 19 | O-$L_1$ | 19.2 | O-$L_1$ |
| 25.5($\pm$0.4) | Bulk plasmon | 24.8($\pm$0.4) | Bulk plasmon | 24.5($\pm$0.4) | Bulk plasmon |
| 33.3 | Na-$L_1$ | 33.5 | Na-$L_1$ | 33.5 | Na-$L_1$ |

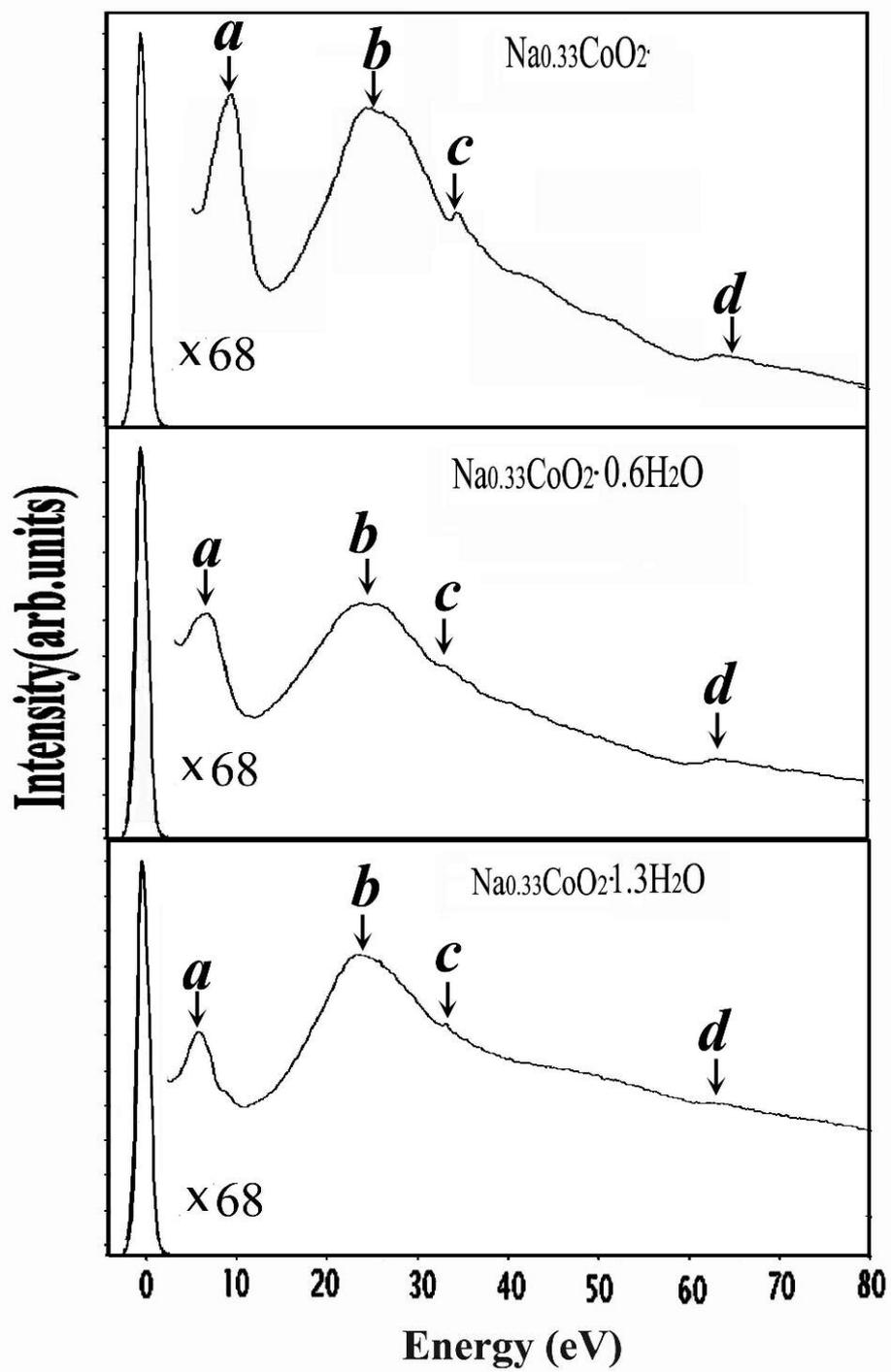

Figure 1

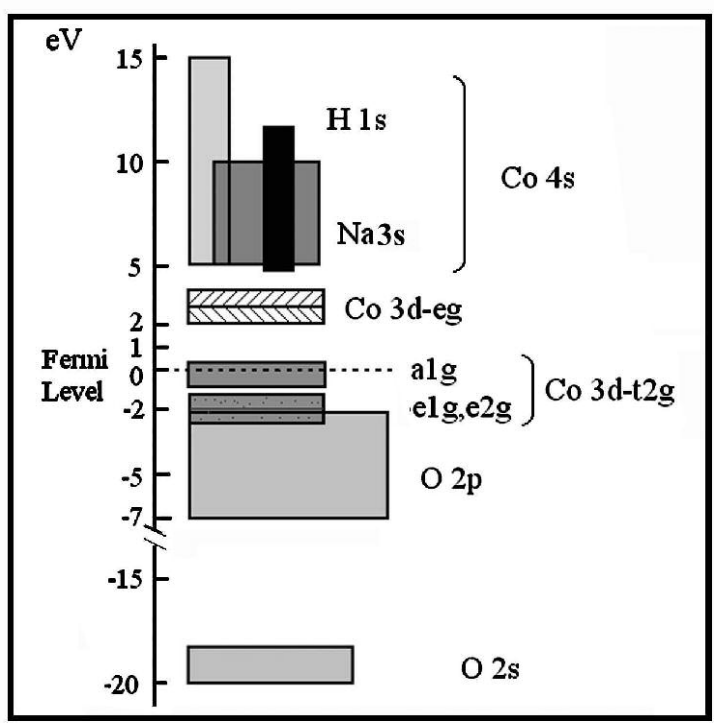

**Figure 2**

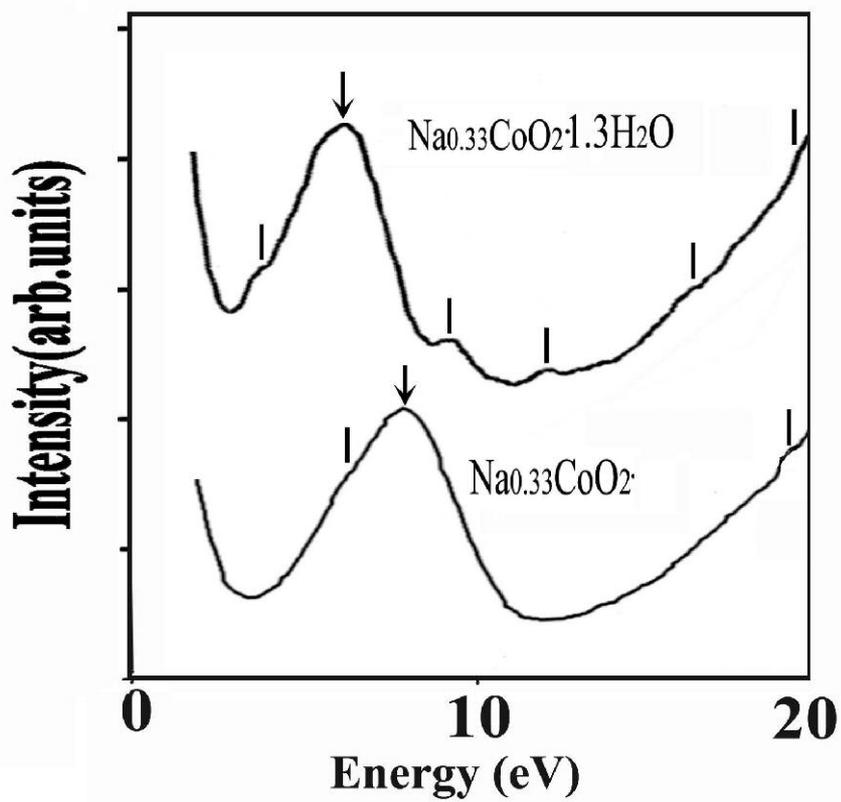

**Figure 3**

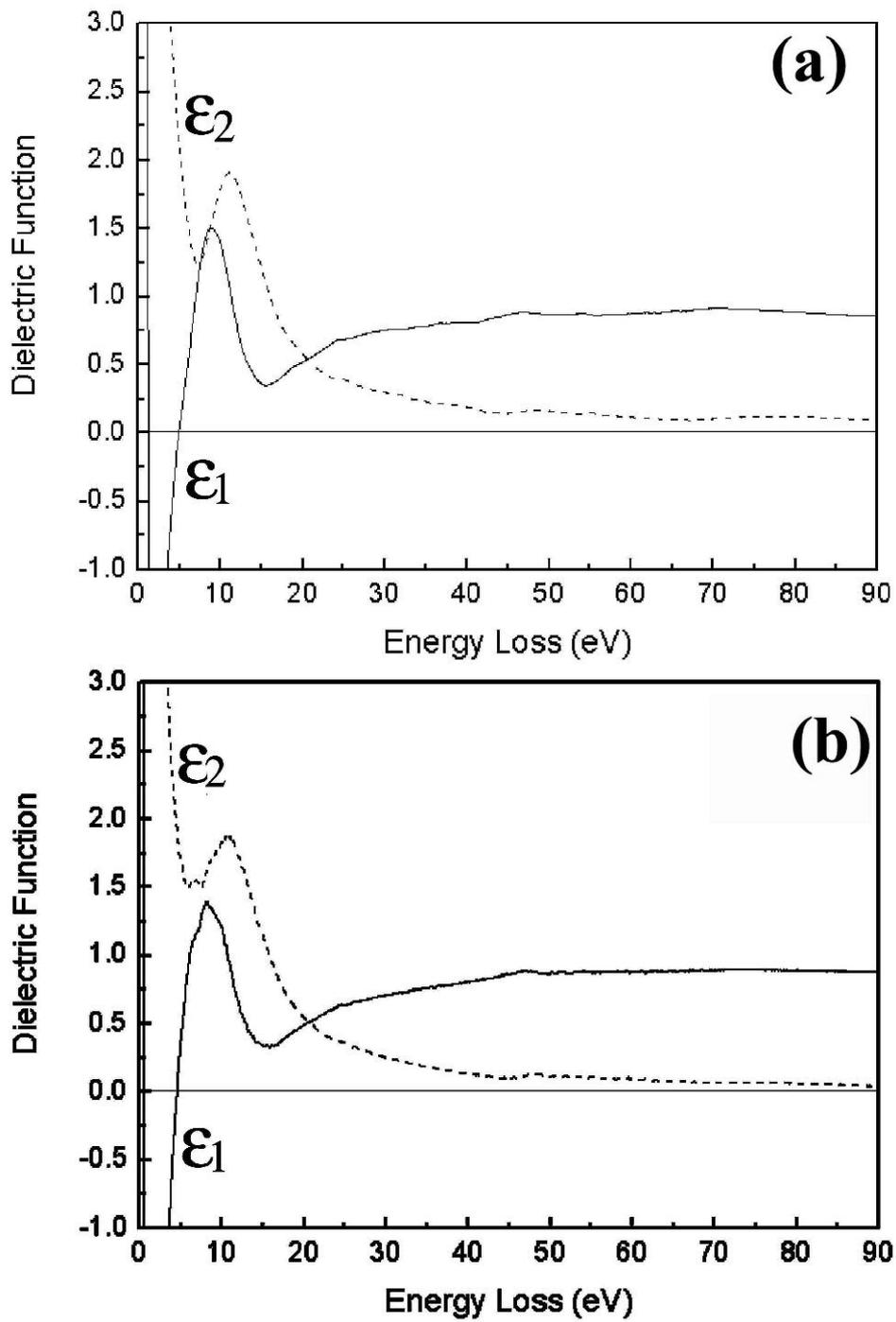

**Figure 4**

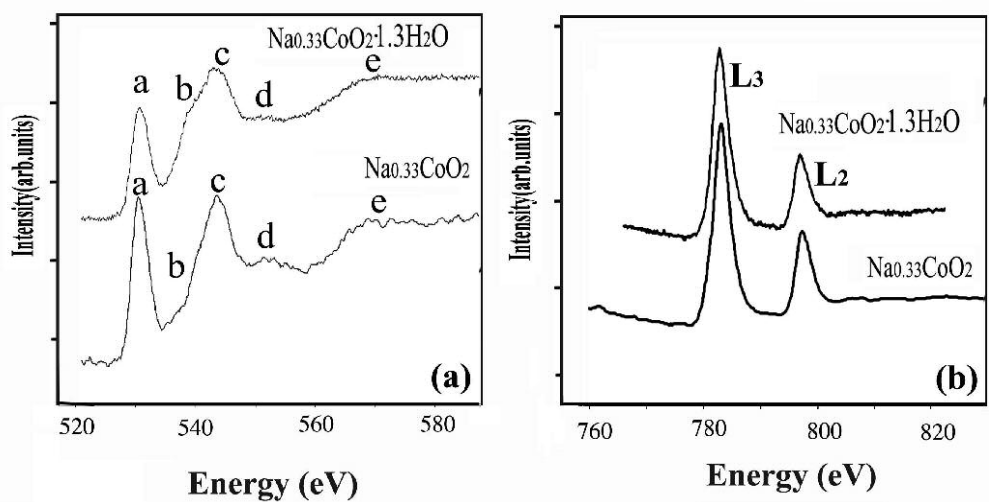

**Figure 5**